\documentstyle[12pt]{article}
\newcommand{\mH}{M_{H}}
\newcommand{\mW}{M_{W}}
\newcommand{\mZ}{M_{Z}}
\newcommand{\mb}{M_{b}}
\newcommand{\mt}{M_{t}}

\newcommand{\pT}{p_{T}}

\newcommand{\signal}{p p \to W H X}

\newcommand{\bgA}{p p \to W b \bar{b} X }
\newcommand{\bgB}{p p \to W Z X }
\newcommand{\bgC}{p p \to W j j X}
\newcommand{\bgD}{p p \to t \bar{t} X }
\newcommand{\bgE}{p p \to t \bar{b} X, \,   \bar{t} b X}
\newcommand{\bgF}{p p \to t \bar{q} X, \,   \bar{t} q X }
\newcommand{\bgG}{p p \to t \bar{b} q X, \,   \bar{t} b q X }

\newcommand{\be}{\begin{equation}}
\newcommand{\ee}{\end{equation}}
\newcommand{\bea}{\begin{eqnarray}}
\newcommand{\eea}{\end{eqnarray}}

\def\agt{\buildrel {\mbox{$>$}} \over {\raisebox{-0.8ex}{\hspace{-0.05in}
$\sim$}}}

\def\alt{\buildrel {\mbox{$<$}} \over {\raisebox{-0.8ex}{\hspace{-0.05in}
$\sim$}}}

\def\mbb{m(b\bar b)}
\def\drbb{\Delta R(b,\bar b)}
\def\dzphibb{\cos \Delta \phi(b,\bar b)}
\def\eORmu{e^{\pm}/\mu^{\pm}}
\def\gamgam{\gamma\gamma}
\def\signature{\eORmu+b\bar b}
\def\zcmHc{z_{\hbox{{\tiny CM}}}^{\hbox{{\tiny RECON}}}(H)}
\def\zcmH{z_{\hbox{{\tiny CM}}}(H)}
\def\zlabH{z_{\hbox{{\tiny LAB}}}(H)}
\def\zlabW{z_{\hbox{{\tiny LAB}}}(W)}
\def\zcmbb{z_{\hbox{{\tiny CM}}}^{\hbox{{\tiny RECON}}}(b,\bar b)}

\def\ie{\hbox{\it i.e.}{}}    
\def\eg{\hbox{\it e.g.}{}}    

\def\etal{\hbox{\it et al.}{}}
\relax

\def\figcap{\section*{Figure Captions\markboth
     {FIGURECAPTIONS}{FIGURECAPTIONS}}\list
     {Fig. \arabic{enumi}:\hfill}{\settowidth\labelwidth{Fig. 999:}
     \leftmargin\labelwidth
     \advance\leftmargin\labelsep\usecounter{enumi}}}
 \relax
\def\reflist{\section*{REFERENCES\markboth
     {REFLIST}{REFLIST}}\list
     {[\arabic{enumi}]\hfill}{\settowidth\labelwidth{[999]}
     \leftmargin\labelwidth
     \advance\leftmargin\labelsep\usecounter{enumi}}}
 \relax
\def\tabcap{\section*{Tables\markboth
     {TABLES}{TABLES}}\list
     {Table \arabic{enumi}:\hfill}{\settowidth\labelwidth{Table 999:}
     \leftmargin\labelwidth
     \advance\leftmargin\labelsep\usecounter{enumi}}}
 \relax

\begin{document}
\begin{titlepage}
 \null
 \vskip 0.5in
\begin{center}
 \makebox[\textwidth][r]{MSU-HEP/40901}
 \makebox[\textwidth][r]{CPP-94-32}

 \vspace{.15in}
  {\Large
  SIGNATURE OF THE  INTERMEDIATE MASS HIGGS BOSON  \\
 \vspace{0.2cm}
     AT THE LHC WITH FLAVOR TAGGING
    }
  \par
 \vskip 1.5em
 {\large
  \begin{tabular}[t]{c}
   Pankaj Agrawal and David Bowser-Chao \\
\em Department of Physics \& Astronomy, Michigan State University \\
 \em East Lansing, Michigan 48824, USA \\
  \\
Kingman Cheung\\
\em Center for Particle Physics, University of Texas at Austin \\
 \em Austin, Texas 78712, USA
  \end{tabular}}
 \par \vskip 5.0em
 {\large\bf Abstract}
\end{center}
\quotation
We revisit the exclusive signature $e^\pm (\mu^\pm) + b\bar b$ in the
associated production of the intermediate mass Higgs boson with a $W$
boson.  We show that it is feasible to use this signature to
identify the Higgs boson in the lower intermediate mass region ($90$
to $130$ GeV) at the LHC with reasonable acceptance cuts by assuming
effective bottom-jet identification.  We further demonstrate that the
background can be reduced with judicious cuts down to the level of the
signal.
\endquotation
\vspace{1.0in}
 \baselineskip 10truept plus 0.2truept minus 0.2truept

\vfill
\mbox{OCTOBER 1994}
\end{titlepage}

\baselineskip=21truept plus 0.2truept minus 0.2truept
\pagestyle{plain}
\pagenumbering{arabic}

\makebox[\textwidth][c]{\bf I. Introduction }

Recently the CDF collaboration presented ``evidence'' for the
existence of the top quark, estimating the top quark mass
to be about 174~GeV [\ref{CDF}].  Should this evidence turn into firm discovery
with the
accumulation of further data, the Higgs boson would become the sole
major missing piece of the  Standard Model. The existence of the Higgs
boson is necessary to lend support to the Higgs sector component of
the  Standard Model, which implements the symmetry breaking mechanism
that keeps the model renormalizable and gives masses to the vector
bosons and fermions.  In the  Standard Model, the Higgs sector consists
of a single scalar Higgs doublet and its interaction with the other
fields of the model.  One can extend the  Standard Model by considering
more than one Higgs doublet.  However,   such extensions usually include a
scalar particle with properties similar to the Higgs boson  of the
Standard
Model. Several non-trivial extensions of the Standard Model, \eg,
supersymmetric models, have such an extended Higgs sector.  At this
moment, there does not exist any satisfactory alternative to the
realization of the Higgs sector by scalar fields. Furthermore,
the detectors at the LHC are being optimized for the detection of
the Standard Model Higgs boson [\ref{ATLASloi}, \ref{CMSloi}].
It is therefore imperative
that the various signatures of the Higgs boson should  be studied in as much
detail as possible. In this paper, we focus on one mode of detection.

 There exists a lower bound on the mass of the Higgs boson of order
$60$ GeV from LEP [\ref{lepbound}].
It is also generally believed that if the mass
of the Higgs boson is of order $1$ TeV or larger, then properties of
the Higgs sector will be quite different.  On the basis of the
production mechanisms and detection strategies, the mass of the Higgs
boson can be classified as light ($\mH < \mZ$), intermediate ($\mZ <
\mH < 2 \mW$), or heavy ($2 \mW < \mH$).  In this
paper, we focus on the intermediate mass region.  For
convenience, we further subdivide this region into two
parts: the lower intermediate mass region ($\mZ < \mH < 130$~GeV) and
the upper intermediate mass region (130~GeV~$ < \mH < 2 \mW$).

In the intermediate mass region, the production mechanism $gg \to H$
gives rise to largest signal rate. Over most of this mass region the
dominant decay mode for the Higgs boson is $H\to b\bar b$.
Combination of these two production and decay mechanisms would
lead to two bottom jets in the final state. Unfortunately, the signal here
is overwhelmed by the background due to the direct production of such
jets  through strong interaction, even if bottom jets could be flavor tagged
[\ref{hunter}]. Therefore, numerous rare decay modes of the Higgs
boson have been examined in the literature.
It has been found that
the decay mode $H \to ZZ^*$ may be useful (where $Z^*$ is an off
mass-shell $Z$ boson) in the upper intermediate mass region
[\ref{gun}]. Due to the necessity of the leptonic decay of both $Z$
bosons, however, with the expected luminosities the number of signal
events will be quite small after experimental cuts are
applied. Also, this scheme will not be useful if $\mH < 130$ GeV,
because of the already small branching ratio for $H \to ZZ^*$ drops
rapidly with $\mH$.

In the lower intermediate mass region, the rare decay mode $H \to
\gamma \gamma$ has been found to be useful. The
signal naturally peaks sharply in the invariant mass $m(\gamma\gamma)$
distribution. If one assumes both very low systematic error and very
high resolution in $m(\gamgam)$, the uncertainty $\sigma$ in the
background, B,  count under the signal, S,  peak is statistical and
approximately equal to $\sqrt{B}$, and the resulting signal
significance $S/\sigma$ is quite high [\ref{GEM}, \ref{SDC}]. Obtaining
low systematic error, however, is far from easy in this case, as the
signal-to-background ratio is only of order ten percent. Due to this
low signal-to-background ratio and the large reducible backgrounds
from mistagging of jets as photons, the shape and the normalization of
the background continuum distribution in $m(\gamgam)$ must be measured
with very high precision to perform the necessary background
subtraction.  Another channel to search for $H\to\gamgam$ is via the
associated production of a Higgs boson with a $W$ boson or a $t\bar t$
pair, with the subsequent decays $H\to \gamma \gamma$ and,
respectively $W \to \ell \nu$ or $t\to b\ell\nu$.   The isolated charged
lepton is then used as an
event trigger to significantly reduce the standard model backgrounds,
resulting in an improved signal-to-background ratio.  At the LHC,
however, the signal rate for these production and subsequent  decay
modes is quite low [\ref{IMH}].

Because of these difficulties in the case of rare decay modes,
the dominant decay channel, $H\to
b\bar b$ has also been examined for the  production
of the Higgs boson in association with a $W$ boson  [\ref{wh}, \ref{ASSW}]
or a $t\bar t$  pair [\ref{ttbb}]. In the case of the Higgs boson production
with a $t\bar t$  pair, one would require tagging of at least three of the
bottom jets. Furthermore,
at LHC energies, the rate for the associated production with a $W$ boson
is higher than production in association with a top quark pair [\ref{hunter}].
We
thus specifically reconsider
\be
 \signal \, .
\ee
The intermediate mass Higgs boson decays predominantly into a $b\bar
b$ pair while the $W$ boson decays into $l^\pm \nu_l$ or two jets.
The decay of the $W$ boson into two jets will result in four jets in
the final state. The rate for such events cannot compete with the QCD
background production of four jets, even assuming perfect bottom jet
identification.  The situation is better if one instead requires the
leptonic decay of the $W$ boson, but the background rate is still
at least two orders of magnitude larger than the signal.  However, by
requiring the two jets from the signal to be initiated by bottom
quarks, we could obtain a significant reduction in the backgrounds.
The problem thus depends critically on bottom-jet identification.

There are several ways to tag heavy quark flavor, of which we shall
briefly discuss two. One technique is to look at the semi-leptonic
decay channel of the quark. Denoting the momentum of the charged
lepton ($\eORmu$) in the plane perpendicular to the motion of the
quark by $p^{T}_{l,q}$, we have $p^{T}_{l,q} < m_{q}/2$.  The
distinguishing feature of this charged lepton, as compared to that
from the decay of the $W$, is that it is {\it within} a jet.  If we
also reject $\eORmu + jj$ events lacking a soft lepton which satisfies
$p^{T}_{l,q} < 1$ GeV, we will be left with a fairly clean event
sample of $\eORmu + b\bar b$. Unfortunately, when we use the
semi-leptonic decay channel of the bottom quark to tag it, there is a
neutrino in its decay products. This means that this method of tagging
will reduce the efficacy of the principal tool for reducing the
background, reconstruction of the Higgs boson, due to decreased
angular and energy resolution of the bottom jets.

Another tagging method is to look for a displaced vertex using a vertex
detector, exploiting the relatively long lifetime of the B
meson. The identification of the secondary vertex structure associated
with the B meson decay is used as a tag, allowing improved
reconstruction of the bottom jets, because the hadronic decay
modes of the bottom quark can also be used.  This method has been used
at $e^-e^+$ colliders [\ref{btag}] and at the Tevatron with success
[\ref{CDF}].
Furthermore,
the ATLAS collaboration at the LHC has included a silicon vertex detector
in their proposal [\ref{ATLASloi}].

These two tagging mechanisms were examined in the context of the
signature $\signature$ in Ref. [\ref{wh}]. There it was found
that this signature may be useful if flavor tagging is done through a
vertex detector.  However, at the time of that study, the top quark
was expected to have a small mass, so the backgrounds from top quark
production were not considered. This signature has recently been
reexamined [\ref{ASSW}], where top quark production was included as a
background. For maximal signal acceptance, the authors imposed only
minimal cuts and assumed tagging to include the leptonic decay modes
of the bottom jets, and thus also assumed a conservatively low energy
resolution for the Higgs boson peak in the $m(b\bar b)$
distribution. While $5\sigma$ significance (taking $\sigma=\sqrt{B}$)
was shown to be possible, the signal-to-background ratios were quite
low, on the order of one part in fifteen for $\mH =100$~GeV. As will be
discussed below, the naive estimate of significance $S/\sqrt B$, where
$S$ and $B$ are the number of signal and background events, does
not take into account systematic errors.  Assuming the systematic
error to be $\epsilon B$, where $B$ is the total number of background
events in a mass-bin of $m(b\bar b)$ around $\mH$, the significance is
more accurately given by $S/\sqrt{(\epsilon B)^2+B}$, where the systematic
and the statistical errors are added in quadrature.  If the systematic
error in the measurement and the background calculation of the
$m(b\bar b)$ is as high as $10\%$, such a low signal-to-background
ratio would mean that the signal would effectively be swamped in
background.

In this paper we take a different tact, emphasizing the necessity of a
much higher signal-to-background ratio for clear detection, taking
into account the possibly dominant systematic error. Based on the
studies by the SDC collaboration [\ref{SDC}], where displaced vertex
tagging of hadronically-decaying bottom jets yielded about $30\%$
efficiency for high $p_T$ bottom jets, we assume energy and angular
resolutions of the bottom jets to be as high as non-bottom jets.
Through the improved resolution and more stringent cuts, we achieve
both higher signal-to-background ratios and significance.
Furthermore, we point out several other cuts that dramatically
increase the signal-to-background ratio. While these cuts may reduce
the naive estimate of significance, the actual signal significance may
increase, depending on the level of systematic error present in the
measurement.

The paper is organized as follows: In the next section, we describe
the various processes that contribute to the total background.  In
Sec. III, we discuss the computation of and the numerical results for the
signal and the various backgrounds. In this way, we assess the
usefulness of the exclusive $\signature$ signature.  In Sec. IV, we
explore ways to enhance the signal-to-background ratio. In Sec. V, we
present our conclusions.

\vspace{0.4cm}

\makebox[\textwidth][c]{\bf II. Background Processes}

\vspace{0.4cm}

    The exclusive signature studied here is $\signature$. By
  exclusive, we mean that we veto any event that has extra hard particles
  other than an isolated charged lepton and two bottom jets.  Therefore,
  any process that can give rise to a $W$ boson and two bottom jets
  with the $W$ boson decaying leptonically is a background.
  Another source of background is due to
  flavor misidentification. A gluon or light quark-initiated jet can
  fake a bottom jet with a small probability; we therefore also
  consider processes that give rise to
potential backgrounds due to this   misidentification.

	The backgrounds include $W$-production in association
with two jets (one, both, or none of which are bottom jets):
\be
\bgA,
\ee
\be
\bgB,
\ee
\be
\bgC.
\ee

The QCD-induced process ($2$) generates a continuum background in
$\mbb$, and is more than an order of magnitude larger than the signal.
Process ($3$) (where the $Z$ boson decays into a pair of bottom
quarks) will be peaked in $\mbb$ around $\mZ$, but due to the
intrinsic width of the $Z$ and the finite resolution in jet energy, it
will be a non-negligible background for Higgs boson masses within
about 10~GeV of $\mZ$. The last process is a background when both jets
are mistagged as bottom jets. Fortunately, the probability of a high
$p_T$ jet faking a bottom jet is quite small, of the order of one
percent. We also note that the jets from process ($4$) tend to have
smaller $\pT$ as compared to the signal, which will help to further
reduce this background.  Calculations of the cross-sections and
distributions for these processes already exist in literature.
   For processes (2), (3) and other backgrounds discussed
below, we have carried out independent calculations, while for process
($4$) we have used an existing program [\ref{ellisprog}].  Since the
calculations are straight-forward, we will not detail our calculations
here.

The other sources of background considered here are top-quark
productions.  Since the D0  collaboration has placed a lower limit of 131~GeV
on the top quark mass [\ref{D0}] --- above the threshold $\mW + \mb$ --- the
top
quark decays with unit probability into a $W$ boson and bottom quark.
Thus the processes listed below, which include production of one or
more top quarks, can also mimic the signal:

\be
\bgD,
\ee
\be
\bgE,
\ee
\be
\bgF,
\ee
\be
\bgG.
\ee

 Strong-interaction process ($5$) is the primary mechanism for the top
quark production at the hadron colliders. It poses as a background to the
electroweak signal in two ways.  The top quark can decay either
semi-leptonically or hadronically, \ie, $t \to b l   \nu_l$ or, $b q
\bar{q'}$.  Since our signature requires one isolated charged lepton,
either the top quark or the anti-top quark must decay semi-leptonically, while
the
other can decay either way. If one top quark decays hadronically and
the other leptonically, there will be two light-quark jets in the
final state in addition to the charged lepton and the bottom-jet pair.
Because the top quarks are very heavy and produced at central
rapidities, the extra jets from the top decay will tend to have large
$p_T$ in the central rapidity region.  Similarly, if both top quarks
decay leptonically the extra lepton will be stiff and central. Either
way, we can reduce the top-pair background by vetoing extra leptons or
jets to as low $p_T$ and as large rapidity as covered by detectors.
Here and
in all our calculations but the $Wjj$ program, we have included full
spin-correlations in the decay of top quarks and $W$ bosons.

Processes ($6$)--($8$) involve electroweak production of a single top
quark. In these processes, the top quark must decay semi-leptonically
to give the charged lepton.  Process ($6$) is, like the signal,
produced through the Drell-Yan mechanism. Despite its strong
similarity to the signal, we shall see in Sec. IV that there do exist
methods to enhance the signal with respect to this background. Process
($7$), which is obtained from the Feynman diagrams of process ($6$) by
crossing, requires a mistag of the light outgoing quark. Finally, the
process (8) mainly comes from gluon-$W$ fusion.  The
extra quark in process ($8$) is mostly produced at forward rapidities,
thus making this background difficult to suppress.  As with the
top-pair background, vetoing extra final state particles (in this
case, the light quark jet) is the most effective cut. In fact, we have
required the signal to be exclusive mainly in order to reduce these
top-production backgrounds.

We have computed the cross-sections for all processes at tree level
for consistency; in any case, the QCD corrections are not available
for all the backgrounds considered.  The question thus arises as to
the effect of these next-to-leading order (NLO) corrections.  A
priori, it is possible that the inclusion of QCD corrections may
require a significant modification in our signature and the sets of
cuts that we consider to achieve the observability of the signal.  The
QCD corrections to the signal production have been calculated, and
found to be about 10--15$\%$ [\ref{willenHan}, \ref{resum}]. In the case of
backgrounds, depending on the process, the corrections could range
from 10--50$\%$ [\ref{nason}].  Therefore, if we were examining an
inclusive signature, we would expect the number of background events
to increase more than that of the signal, which could lead to a
decrease in the significance.  However, we will argue that higher
order QCD corrections will not significantly affect the kinematics of
the final state particles for the signal and backgrounds in our
analysis.  Therefore, our set of judicious cuts should also be
applicable when higher QCD corrections are taken into account.

Since we are examining an exclusive signature, the issue of initial
state radiation is especially important. As remarked above, the
exclusive nature of the signal serves to cut down two major
backgrounds: $\bgD$ and $\bgG$.  In Sect. IV, we impose an extra
lepton/jet veto to the maximum coverage allowed by the detector,
\ie,  the minimum $p_T(\ell/j)$ and maximum pseudo-rapidity
$|\eta(\ell/j)|$. At tree level, this cut has no apparent effect on
the signal; on the other hand, Altarelli-Parisi evolution of the
parton distribution functions implies that the hard scattering process
will necessarily be accompanied by extra jets, whose energy scale
should be small compared to the hard scattering energy scale. Because
this scale is of order $\mH+\mW \approx 200$~GeV, we expect a
significant fraction of the signal and background events to include
one or more extra jets of $p_T \agt 15$~GeV.

As is well-known, the NLO calculation is an unreliable measure of the
frequency of this extra radiation; we expect an estimate based on
fixed-order perturbation theory to break down below a certain value of
$\pT$ for the extra jet. Resummation indicates where this breakdown
takes place; for the signal at the LHC, one expects this minimum $p_T$
to be around 25--30~GeV [\ref{resum}].  Resummation, in turn, does not
offer any information about the number of extra soft jets or their
individual $p_T$; so to help answer these questions, we have employed
a Monte-Carlo generator (PYTHIA [\ref{pythia}]).

This type of Monte-Carlo simulation usually includes only
bremsstrahlung-type corrections; inclusion of loop-corrections in a
consistent manner has been implemented in some specialized Monte-Carlo
generators in the case of a very few processes. However, our interest
is in that part of NLO corrections that gives rise to extra soft jets,
for which purpose PYTHIA can serve fairly well (though we note that the total
cross-sections calculated by such programs are normalized to LO only, thus
necessitating multiplying results by an overall K-factor).

To show that our analysis would not be significantly altered by the
inclusion of NLO corrections, we have carried out a simple study using
PYTHIA.  The results discussed at the end of the next section
essentially imply that our veto of extra jets may be modified slightly
so as to allow for about the same signal event rate as obtained with
our cuts on the NLO cross-sections, while sufficiently suppressing
processes~($5$) and~($8$).

\vspace{0.4cm}

\makebox[\textwidth][c]{\bf III. Numerical Results}

\vspace{0.4cm}

	 In this Section, we present the signal and background event
rates at the LHC. Because the collider and its detectors are not
expected to become operational for almost a decade, actual detector
resolution and efficiency are of course still uncertain; for our study,
we have employed a simple detector simulation based on the
parameters presented in the ATLAS collaboration's Letter Of Intent
(LOI) [\ref{ATLASloi}].

To bracket the range of detector performance of  the ATLAS collaboration,
we consider two scenarios. The first scenario is for more optimistic detector
performance, while the second scenario assumes more conservative
detector acceptance.  In addition, the differences between the two
scenarios include the efficiency assumed for tagging the $b$-quarks
and in the veto for extra final state particles.  Explicit details are
summarized in Table~1.  We assume the same hadronic and lepton
Gaussian energy resolutions for both scenarios (also given in the
table and taken from the ATLAS LOI). As part of the cuts outlined
below, we veto an event if an extra electron or muon is detected in
the central region; in the forward region ($3<|\eta|<4.5$), we rely on
the hadronic calorimeter to detect and veto electrons (but not muons).
No angular resolution effects have been included. The missing
transverse energy $\not{\!\!E_T}$ is calculated by adding up all the
visible (smeared) momenta and assuming hermetic coverage of soft
forward radiation.

Because of large hadronic backgrounds at the LHC and because our
signature involves two bottom jets, it is clear that $b$-tagging is
crucial.  B-tagging is characterized both by true $b$-jet tagging
efficiency and light quark/gluon mistagging rates.  Studies by the SDC
[\ref{SDC}] and ATLAS [\ref{ATLASloi}] collaborations estimated $b$-tagging
through displaced vertices to have an efficiency of about $30\%$ and a
mistagging rate of $1\%$ for jets of $p_T \agt 40$~GeV,
with degraded $b$-tagging efficiency for lower $p_T$ jets due
to the correspondent decrease in impact parameter resolution. We have
taken these efficiencies to be independent of $p_T$, but for the more
conservative scenario~(ii), we require a higher minimum $p_T$ than in
the more optimistic scenario~(i). Finally, we take the energy
resolution of the bottom jets
to be that of ordinary jets, because we are relying on
displaced vertices for $b$-tagging instead of soft, non-isolated
leptons and their concomitant neutrino(s). Furthermore, since the
fragmentation of the bottom jets is expected to be harder than for
non-bottom jets, one would expect bottom jet energy resolution to  be
somewhat better than for light quark or gluon jets.

Table~1 shows that the acceptance cuts for the tagged lepton
are tighter in the second scenario than the first one.
The larger pseudo-rapidity coverage in scenario~(i) is not
unreasonable, however, assuming that tagging of either of the two
bottom jets can serve as efficient as the lepton tagging for an event
trigger. With regard to the minimum $p_T$, the LOI indicates that
coverage as low as 6--10~GeV should be considered;
if a more detailed study showed sizeable lepton backgrounds from heavy
flavor decay, one could employ the more stringent cuts of scenario~(ii).
As explained above, the transverse mass cut is imposed to ensure
consistency with the presence of a $W$-boson. For the purpose of this
cut, in both scenarios we have included forward ($|\eta|>3$) muons as
a source of  $\not{\!\!E_T}$.

The most significant difference between the two sets of cuts is the
treatment of vetoing extra particles.  According to the ATLAS LOI, the
rapidity coverage of the hadronic calorimeter is expected to be up to
$5$. The central issue for forward jet/$e$ vetoes is thus to what
minimum $p_T$ value these jets will be visible. We emphasize that we
are not assuming precision measurement of forward, soft jets, since
these jets are not part of our signal. Any observation of a forward
jet would suffice, as long as the underlying event for each signal
event did not mimic such jets very frequently. In any case, in both
scenarios we assume $100\%$ efficiency for tagging jets/electrons of
$p_T > 15$ GeV within the active region of the calorimeter
($|\eta|<4.5$), and for scenario~(ii) we take a reduced efficiency of
$50\%$ for jets/electrons with $10\; \hbox{GeV}<p_T<15$ GeV in the same
rapidity range.

Before presenting results for the two scenarios, we briefly describe
the inputs to our calculation.
  First, we have conservatively assumed a yearly LHC integrated
luminosity of 10 fb$^{-1}$. Although the luminosity is eventually
expected to grow by a factor of 10--50 larger, it is quite possible
that only the nominal luminosity will be attainable in the first few
years of operation. Furthermore, while the rate for overlapping
$b$-tagged events (with accompanying tagged leptons) at a much higher
luminosity is not expected to pose a problem, it is not clear
whether a silicon vertex detector would be able to survive such an
environment.
Because all the processes were calculated here to leading order, we
employed the CTEQ2 set $5$ distributions  [\ref{cteq}], which are leading order
fits. Similarly, $\alpha_s$ is calculated to leading order with the value
of $\Lambda_{\rm QCD}$ given by the parton distributions. The
factorization/renormization scales were selected as follows: for the
two s-channel processes ($WH$ and $tb$ production), we took $Q^2 =
\hat s$. For the others, which were at least in part t-channel processes,
we took $Q^2 = \hat s/4$. Though the cross-sections demonstrate
dependence on the choice of this scale (most notably process~(5)),
only if all the next-to-leading order calculations are available could
the scale-dependence be reduced.  In their absence, we have used a
representative scale for each of the processes, taking into account
the nature of the exchange involved. However, reasonable variations in the
factorization/renormalization scale are not expected to significantly
alter the results for the significance and the signal-to-background
ratio.

We present our results for the two scenarios in Tables~2--5. As can be
seen, in both scenarios we can achieve very good statistical
significance $S/\sqrt{B}$ (where we for the moment ignore systematic
errors). One would therefore expect the signature under consideration
to serve quite well as a detection mode. For $ \mt = 175$ GeV, for
example, over most of lower intermediate mass range one can obtain a
significance of four or higher for 10 fb$^{-1}$ integrated luminosity.
For $\mH\approx 90$ GeV the worst backgrounds are $Wb\bar b$ and $WZ$, the
latter of which drops substantially as $\mH$ increases from 90 GeV.
For the other Higgs boson masses, the largest backgrounds are $Wb\bar b$
and $t\bar t$ for smaller top mass $\mt \alt 175$ GeV, but for larger top
mass $\mt \agt 175$ GeV $tbq$ and $Wb\bar b$ backgrounds are the most serious.
Through our strategy of requiring an exclusive signature, we have
succeeded in reducing the top-pair and $ tqb$ backgrounds to an
acceptable level.  A fairly high minimum $p_T$ for the bottom jets has
also worked rather well in decreasing the $Wjj$ background relative to
the signal, as can be seen from comparing the results of the two
scenarios. This is expected as the jets in $Wjj$ are mainly
bremsstrahlung gluons off quark or gluon lines.

The cuts of the two scenarios were selected for their effect on the
significance ignoring the signal-to-background ratio. In the next
section, we shall discuss strategies to help in reducing the signal to
background ratio, which would be crucial in the event that systematic
errors dominate the background uncertainty.

We now explore the consequences of considering the NLO corrections to
the cross-sections and the kinematics of the signal and background
events.  As discussed in Sec. II, we do so through a simple simulation
employing PYTHIA.  The focus was on investigating the necessary
modifications to our simple ``extra jet/lepton'' veto that would
retain a large fraction of the signal while still sufficiently
suppress the $t\bar t$ and $tbq$ backgrounds.

We included initial state radiation (but no final state radiation or
hadronization) to study the differences between initial state
radiation in the signal and in the backgrounds. Setting the
threshold for extra jets at $\pT > 15$~GeV, we find that about $40\%$
of the signal events have one or more extra jet in the central
rapidity region, with the bulk of such events containing only one
extra jet.  Thus our simple veto of as few as one extra jet would cut by
more than a third the signal acceptance (including the NLO
normalization).  The top-pair background was accompanied by extra jets
about 60--70$\%$ of the time.  For the other background processes we
expect similar rates for extra jets.

Allowing up to one extra jet, and vetoing an event only if it had {\sl two}
or more extra jets would yield a higher signal acceptance, help to
reduce the high jet-multiplicity top pair background, but leave the
$tbq$ background untouched at the LO.  Fortunately, it turns out that the extra
jet in the case of the signal tends to be soft --- to pass the minimal
$\pT$ cut of 15~GeV, the jet is quite central and with an energy not
much larger than its transverse momentum. For $\bgG$, in contrast, the
extra jet to be vetoed has both large $\pT$ and large energy. By
vetoing events with a single extra jet having either $\pT > 50$~GeV, or $E >
150$~GeV, this background is suppressed to about the same extent as in
scenario~(ii), with signal acceptance on par with our LO estimates and
cuts.  The top-pair background tends to be accompanied by three or
more extra jets, so even this relaxed jet veto will drastically reduce
this background. We estimate that inclusion of events with one extra
jets would enhance the top pair background at most by 50--75$\%$,
which represents less than 10$\%$ increase in the total background rate.


Finally, we emphasize that though this preliminary study indicates
that (with the inclusion of the NLO corrections) our modified cuts
will likely lead to comparable signal acceptance and
signal-to-background ratio, a full study of all the processes, including
initial
and final state radiations, will be a necessary check of these results.

\vspace{0.4cm}

\makebox[\textwidth][c]{\bf  IV. Enhancing Signal-To-Background}

\vspace{0.4cm}

 The cuts presented in the previous section were of two types:
acceptance cuts and significance-enhancing cuts. Acceptance cuts were
just those made to ensure the visibility of the signal, regardless of
the background acceptance, such as requiring minimum transverse
momenta for the lepton and bottom quarks. Significance-enhancing cuts
included the extra lepton and jet vetoes, and the requirement that the
bottom-quark pair closely reconstruct the Higgs boson. These cuts were
intended to maximize the naive estimate of the signal significance,
$S/\sqrt B$, where we ignored the effect of systematic error.

In practice, of course, systematic error could be relatively
important. We have performed a background subtraction to obtain the
various values for the significance given in the previous section, where we
assumed
accurate measurement of the background over some reasonable range in
$m(b\bar b)$. Aside from the theoretical uncertainty in the various
backgrounds, the relatively low event rates for the background (with
all but the Higgs boson reconstruction cut applied) and the difficulty
in reconstructing bottom jets would certainly imply some systematic
error in this measurement.  Given that the highest signal/background
ratio in scenario~(ii) was only about one part in four,  as discussed
in Sec. I, it is important to seek additional cuts that can
increase the signal-to-background ratio,
which in practice could be the true figure of merit.  In this section,
we examine several such cuts. For the sake of simplicity, we will
present results for typical top-quark and Higgs-boson masses at $\mt =
$175~GeV, $\mH = $100~GeV, and starting with the cuts in scenario
(ii).

As with most of the
rest of this study, we are considering leading order effects only ---
radiative corrections can affect the practical efficacy of
these cuts. The physical differences between the signal and various
backgrounds are, however, most apparent at the leading order, and our
results should suggest analogous cuts for more detailed future studies
that include initial and final state radiations, b-fragmentation, etc.

  From Tables $2$ and $4$, we see that direct $W$ and top production each
make up significant portion of the total background. Considering first the $W$
production processes, we have found that the spatial separation between the
bottom quark and antiquark, measured
in various ways, to be quite useful. As previously reported
[\ref{proceed}], for example, the variable $\drbb$ defined by
$\Delta R(b,\bar b)=\sqrt{(\Delta \eta)^2 + (\Delta \phi)^2}$, where
$\Delta \eta$ is the difference in pseudorapidity and $\Delta \phi$
is the difference in azimuthal angle between $b$ and $\bar b$,
is very useful in
reducing these backgrounds. The signal is roughly flat in the range $1
< \drbb < \pi$, while the $W$ production backgrounds peak towards
$\drbb=\pi$.
It turns out that this cut may be replaced by a simpler one in
$\dzphibb$, which is the cosine of the azimuthal angle between $b$ and
$\bar b$.  Figure~1 shows a clear concentration of the $W$ backgrounds
towards $\dzphibb=-1$,\ie, where the jets are back-to-back in the
transverse plane. Though not shown separately, the $Wjj$ and $Wb\bar b$ are
responsible for this difference from the signal; the $WZ$ background
is much  more
similar to the signal. This behavior reflects two characteristics of these
backgrounds. Because the $Wjj$ and $Wb\bar b$ backgrounds tend to have low
$\mbb$, it is easiest for them to reconstruct the Higgs boson with the
minimum of energy, by producing the $b\bar b$ pair back-to-back.
Due to the t-channel enhancement (of all three processes),
in the center-of-mass
frame of the $W+b\bar b$ system, the bottom quark pair is produced
predominantly in the forward/backward regions. In the laboratory
 frame, the
combination of these effects manifests itself as a strong peak at
$\dzphibb = -1$.

 Unfortunately, the backgrounds from top quark production do not
differ significantly from the signal in this distribution. Imposing the
severe cut of $\dzphibb > 0.5$ (Table~6) increases $S/B$ by a factor
of 1.5, though at the cost of keeping only about $15\%$ of the
signal. A looser cut would not appreciably increase $S/B$, because of
the top quark backgrounds. Provided one is willing to try the
center-of-mass (CM) frame reconstruction, however, we show below that
$\zcmbb$ (the cosine of the angle between $b$ and $\bar b$
quarks in the CM frame) can help to
reduce the top backgrounds quite significantly as well. Here and
below by the CM frame we mean the CM frame for the
$W+b\bar b$ system.

Another reason to consider CM frame reconstruction is the strong potential of
the observable $\zcmH$, which is the cosine of the polar angle of the
Higgs boson in the $W+b\bar b$ center-of-mass frame.
As discussed above, the t-channel enhanced $W+X$
backgrounds are strongly peaked in the forward/backward region,
while the signal is centrally peaked in the CM frame (see Fig.~2).

The  angular distribution $\zlabH$ of the Higgs boson in the
laboratory frame, shown in Fig.~3, is clearly much less
informative, due to  the washout by the longitudinal  boost of the
incoming partons.
{}From Fig. 2 it is clear that a cut on $\zcmH$ can greatly reduce the $W+X$
backgrounds.  Here, the challenge is to best reconstruct the
leptonically-decaying $W$ boson, and then the CM reference frame,
despite the indirect detection of the neutrino, the intrinsic width of
the $W$ boson, and limited detector resolution.  The principal
difficulty in reconstruction stems from having to choose between the
two solutions of the quadratic equation for the neutrino longitudinal
momentum.

We have not attempted to find the optimal reconstruction algorithm,
which we leave to more detailed studies. To merely point out the
potential of our suggested CM frame cuts, however, we have
selected the solution for the $W$ boson lying closest in $\Delta R$
to the reconstructed Higgs boson. As is shown below, even this simple
choice yields remarkable enhancement of $S/B$, which should motivate
more detailed study of the issue.

Thus reconstructing the CM frame, we obtain the observable $\zcmHc$,
which approximates $\zcmH$, shown in Fig.~4 for the signal and
backgrounds. Comparing to the cut in $\dzphibb$, a cut in this
variable manages to gain about the same increase in $S/B$, but at
about three times as much in signal acceptance as before (Table~6).
Further
improvement is possible by including a cut on the total invariant
mass, $M_{Wb\bar b}^2 = (p_W+p_{b\bar b})^2$. Due to the fact that
 requiring a larger invariant mass $M_{Wb\bar b}$
 somewhat improves the accuracy of our simple reconstruction
algorithm, sharpening the $W+X$ forward/backward peak, and leaving
proportionally more of the signal in the central region.  With these
two additional cuts (results given in Table~6), we have more than
doubled the ratio of signal to $W+X$ backgrounds.  The $\zcmHc$ cut is
somewhat less efficient with respect to the top backgrounds, a fact
which conversely may be useful in top quark {\sl detection} above
$W+X$ backgrounds [\ref{toppaper}]. Finally, we note that even without
direct reconstruction of the $W+b\bar b$ rest frame, a less efficient
version of this cut is possible by considering the {\sl lab frame}
double distribution in ($\zlabH,\zlabW$) in which  the $W+X$
background has peaks at  (1,1) and (-1,-1) but not for  the signal.

One can also differentiate between the signal and the backgrounds
using $\zcmbb$, which is the cosine of the angle between the two
bottom quarks in the CM frame of $W+b\bar b$.
Figure~5 and Table~6 show the efficacy
of this observable in reducing the top backgrounds but less effective
in reducing the $W$ production backgrounds.   This effect may
be understood for the single-top quark processes, because in the CM
frame the two bottom quarks tend to be produced in opposite
hemispheres with respect to the plane transverse to the beam
direction.  Because this cut is somewhat orthogonal to that in
$\zcmHc$, we have included the effects of applying both types of cuts in
Table~6, which shows an overall increase in $S/B$ by a factor of four.
Finally, we achieve a  signal-to-background ratio close to 1:1.

Assuming the ability to reconstruct the $W$ boson, it is natural to
consider top-reconstruction vetoes to reduce the top-quark backgrounds.
Reconstructing the $W$ boson as described above, we have found that
such a cut is possible, but due to the $W$ boson and top quark widths, and
especially detector resolutions, it is difficult to significantly
increase the signal-to-background ratio.

Finally, Figs.~6 and 7 shows that the bottom jets from the signal have
relatively
high  $p_T$ and energy, a fact which may be important in $m(b,\bar
b)$ resolution. In this study, we have assumed the displaced-vertex
tagging only of hadronically decaying bottom quarks, because the energy
resolution of the leptonically decaying bottom quarks is degraded due
to the presence of extra neutrinos. Directly vetoing this decay mode
is difficult, so that if necessary, requiring the jets to have higher
energy and transverse momenta will help with angular and energy
resolution of the jets, both by forcing the neutrino(s) to travel in the
direction of the jet and also to soften leptonically-decaying bottom jets.

\vspace{0.4cm}

\makebox[\textwidth][c]{\bf V. Discussion and Conclusions}

\vspace{0.4cm}

    In this paper, we have studied the exclusive signature
 $\signature$ for the associated production of the Higgs boson with
 mass in the lower intermediate mass region. Such a signature results
 through the process $pp\to WHX \to \ell\nu_{\ell} b\bar bX$.  We
 considered the principal backgrounds and found that one can achieve
 fairly good observability for the signal. We have also studied a few
 strategies that one could adopt to enhance the signal-to-background
 ratio. For the sake of consistency, all processes were studied at the
 LO; however, a brief study was carried out using PYTHIA to find the
 implications of NLO corrections.

We have studied two scenarios: optimistic and conservative.  In both
scenarios, we achieve fairly good statistical significance for the
signal, reasonable signal-to-background ratio and sufficient number
of signal events.  However, we also argue that due to potentially serious
systematic errors it is advantageous to devise special cuts to further
suppress the backgrounds.  Given the very different nature of
the production mechanisms for the signal and various backgrounds,
it is natural to expect that wide variety of observables exist that
could help us in improving the signal-to-background ratio.
We found that the observables  $\drbb$ and $\dzphibb$ in the
laboratory frame, and the observables $\zcmHc$
and $\zcmbb$ in the CM frame of the $W$ boson and the $b\bar b$ pair,
or a combination of them are very useful in enhancing the
signal-to-background ratio.  Finally, we achieved the
signal-to-background ratio close to 1 with about 10 signal events for
10 fb$^{-1}$ integrated luminosity.

 Our goal was not to find the best set of cuts; this is best left for
 a study that includes detailed detector simulation. What we have
 shown is that the observation of the considered signature (or a
 variation) can lead to the detection of the Higgs boson if such a
 Higgs boson in the appropriate mass range exists.  The variation of
 the signature that was briefly considered included the $\signature$
 events with {\em or} without one (sufficiently soft) extra jet, with
preliminary results obtained using PYTHIA.

  The values of the statistical significance that we found are listed
 in the Tables $3$ and $5$. These results will certainly be degraded
 if the bottom-tagging efficiency or jet energy resolution are worse
 than our assumptions, which were motivated by the LOI of the ATLAS
 collaboration and the technical design report of the SDC
 collaboration. On the other hand, the integrated luminosity may well
 be higher than what we have taken as a value, with obvious
 improvements in significance. Beyond collider and detector
 parameters, further improvements are possible in the cuts described
 above. Because in general, strong correlations exist between such
 observables, one might suspect that optimal cuts should be placed in
 higher-dimensional scatter plots. Such tools as neural networks or
 decision trees [\ref{bowchao}] have obtained signal-to-background
 enhancements superior to conventional cuts.

In this study, we only use the electronic and muonic decays of the $W$ boson.
We have not considered the contribution of $\tau$ decay of the $W$
boson and subsequent decay of the $\tau$ into electron or muon.  Such
decays of the $W$ boson will contribute to both the signal and the
background. However, such contributions are expected to be small
due to the inclusion of the further branching ratios of the
$\tau$ into electron or muon.  Furthermore, these
secondary electrons and muons are less energetic than the direct leptons
from $W$ decays, and therefore will have a lower acceptance rate.

So far, we have discussed only the Standard Model Higgs boson.  In extensions
of the Standard Model that have a larger Higgs sector, this mode can
only be used in searching for the lightest neutral Higgs boson.  The
detectability of the lightest Higgs boson, using this associated
production with a $W$ boson and the decay of $Wh\to \ell\nu b\bar b$,
depends on the $WWh$ and $hb\bar b$ couplings, which in turns depend
on the parameter of the extended Higgs sector.  For example, in the
minimal supersymmetric Standard Model, the $WWh$ coupling is in
general reduced from its Standard Model value, while the $hb\bar b$ is
enhanced relative to the Standard Model value.  So it remains possible
to observe the lightest Higgs boson of the minimal supersymetric
Standard Model, but it will depend very much on the parameters of the
model.

Finally, our focus was on the search of the Higgs boson at the LHC.
For a search of the Higgs boson using the signature $ \signature$ at
the Tevatron, the major backgrounds to the signal are due to the $Wb
\bar b$ and $Wjj$ production. Our discussion of the Sec.  IV points
out a few ways to reduce these backgrounds, \eg, by applying cuts on
$\drbb$, $\dzphibb$, or $\zcmH$ observables.  Using decision tree, one
can use these or some other observables to find ways to enhance the
signal with respect to the backgrounds [\ref{whpaper}].

\bigskip
\bigskip
\bigskip

\noindent{\large ACKNOWLEDGEMENTS}

\medskip

We are indebted to Duane A. Dicus for early participation and input.
We would also wish to thank the following for useful discussions:
H. Weerts, B. Pope, C. P. Yuan, J. Pumplin, G. Ladinsky, S. Willenbrock,
and A. Stange. P.A. was supported in part by a NSF grant number
PHY-9396022.
 D.B.C. was supported in part by a NSF grant number PHY-9307980.
K.C. was supported in part by a DOE grant number DOE-ER-40757-056.

\vskip .5in

\relax
\def\pl#1#2#3{
     {\it Phys.~Lett.~}{\bf B#1} (19#3) #2}
\def\zp#1#2#3{
     {\it Zeit.~Phys.~}{\bf C#1} (19#3) #2}
\def\prl#1#2#3{
     {\it Phys.~Rev.~Lett.~}{\bf #1} (19#3) #2}
\def\rmp#1#2#3{
     {\it Rev.~Mod.~Phys.~}{\bf #1} (19#3) #2}
\def\prep#1#2#3{
     {\it Phys.~Rep.~}{\bf #1} (19#3) #2}
\def\pr#1#2#3{
     {\it Phys.~Rev.~ }{\bf D#1} (19#3) #2}
\def\np#1#2#3{
     {\it Nucl.~Phys.~}{\bf B#1} (19#3) #2}
\def\ib#1#2#3{
     {\it ibid.~}{\bf #1} (19#3) #2}
\def\nat#1#2#3{
     {\it Nature (London) }{\bf #1} (19#3) #2}
\def\ap#1#2#3{
     {\it Ann.~Phys.~(NY) }{\bf #1} (19#3) #2}
\def\sj#1#2#3{
     {\it Sov.~J.~Nucl.~Phys.~}{\bf #1} (19#3) #2}
\def\ar#1#2#3{
     {\it Ann.~Rev.~Nucl.~Part.~Sci.~}{\bf #1} (19#3) #2}
\def\ijmp#1#2#3{
     {\it Int.~J.~Mod.~Phys.~}{\bf #1} (19#3) #2}
\def\cpc#1#2#3{
     {\it Computer Physics Commun. }{\bf #1} (19#3) #2}
\begin{reflist}

\item \label{CDF}F. Abe {\it et al.} (CDF Collaboration), \pr{50}{2966}{94}.

\item \label{ATLASloi} ATLAS Letter of Intent, CERN/LHCC  92-4 (October 1992).

\item \label{CMSloi} CMS Letter of Intent, CERN/LHCC 92-3 (October 1992).

\item \label{lepbound} E. Gross and P. Yepes, \ijmp{8}{407}{93}.

\item \label{hunter} J.F.~Gunion \etal, The Higgs Hunter's Guide,
Addison-Wesley  (1990).

\item \label{gun} J.F.~Gunion, G.L.~Kane, and J.~Wudka, \np{299}{231}{88};
V. Barger, G. Bhattacharya, T. Han, and B.A. Kniehl,  \pr{43}{779}{91}.

\item \label{GEM} GEM Technical Design Report, GEM-TN-93-262 (April
1993).

\item \label{SDC} SDC Technical Design Report, SDC-92-201 (April 1992).

\item \label{IMH} Z. Kunszt, Z. Trocsanyi, and W.J. Stirling,
\pl{271}{247}{91};
W. Marciano and F. Paige,\prl{66}{2433}{91};
W.J. Stirling and D.J. Summers, \pl{283}{411}{92}.

\item \label{wh} P. Agrawal and S. Ellis, \pl{229}{145}{89}.

\item \label{ASSW} A. Stange, W. Marciano, and S. Willenbrock,
\pr{49}{1354 }{94};  preprint ILL-TH-94-8 (April 1994) and references there in.

\item \label{ttbb}T. Garavaglia, W. Kwong, and D. Wu, \pr{48}{R1899}{93};
J. Dai, J. Gunion, and R. Vega, \prl{71}{2699}{93}.


\item \label{btag} See, \eg, R. Akers \etal  (OPAL Collaboration),
\zp{61}{209}{94}.

\item \label{ellisprog} S. D. Ellis, R.~Kleiss, and W.J.~Stirling,
\pl{154}{435}{85}.

\item \label{D0} S. Abachi {\it et al.} (D0 Collaboration), \prl{72}{2138}{94}.

\item \label{willenHan} T. Han and S. Willenbrock, \pl{273}{167}{91};
J. Ohnemus and W.J. Stirling, \pr{47}{2722}{93};
 H. Baer, B. Bailey, and J.F. Owens,\pr{47}{2730}{93}.

\item \label{resum} P. Agrawal, J. Qiu, and C.  P. Yuan, MSU-HEP-93-8;
P. Agrawal, J. Qiu, and C.  P. Yuan, Proceedings of the Workshop on
{\em Physics at Current Accelerators and Supercolliders}, Argonne
National Laboratory (June 2--5, 1993).

\item \label{nason}P. Nason, S. Dawson, and R.K. Ellis,  \np{303}{607}{88};
W. Beenakker, H. Kuijf, W. L. Neerven and J. Smith, \pr{40}{54}{89}.

\item \label{pythia} T. Sj\"{o}strand, \cpc{39}{347}{86};
T. Sj\"{o}strand and M. Bengtsson, \cpc{43}{367}{87}.

\item \label{cteq} J. Botts \etal  (CTEQ Collaboration), \pl{304}{159}{93}.


\item \label{proceed} P. Agrawal, D. Bowser-Chao, K. Cheung, and D. Dicus,
preprint MSU-HEP/40815, to appear in Proceedings of DPF'94, Albuquerque,
New Mexico (August 1994).

\item \label{toppaper} P. Agrawal,  D. Bowser-Chao,  and J. Pumplin,
preprint  MSU-HEP/41101.

\item \label{bowchao} D. Bowser-Chao and D. Dzialo,\pr{47} {1900}{93}.

\item \label{whpaper} P. Agrawal,  D. Bowser-Chao,  and J. Hughes,
preprint  MSU-HEP/41102.

\end{reflist}

\newpage


\begin{tabular}{||c|c|c||} \hline
  &  \multicolumn{2}{c||} {Scenarios} \\ \cline{2-3}
  Acceptance and Cut Parameters  & (i) Optimistic & (ii)
Conservative \\ \hline
%
Energy Resolution ($\Delta E/E$): & \multicolumn{2}{c||}{}\\
central $e/\mu$ ($|\eta| < 3$) &
	\multicolumn{2}{c||} { $10/\sqrt{E}\%\oplus 1\%$} \\
central jets ($|\eta| < 3$) &
	\multicolumn{2}{c||} { $50/\sqrt{E}\%\oplus 3\%$} \\
forward  jet/$e$ (3 $< |\eta | <4.5$) &
 	\multicolumn{2}{c||} { $100/\sqrt{E}\%\oplus 7\%$} \\ \hline
%
%
$\max |\eta(b)|$ &  2 &  2 \\
$\min p_T(b)$& 20 GeV & 30 GeV \\   \cline{2-3}
$\min \Delta R (b,\bar b)$ &  \multicolumn{2}{c||}{ 0.6 } \\
$b$-tagging efficiency & \multicolumn{2}{c||} {$30\%$} \\
non-$b$ jet acceptance & \multicolumn{2}{c||} {$1\%$} \\ \hline
Higgs reconstruction
	& \multicolumn{2}{c||}{$|m(b\bar b) - \mH| < $7.5 GeV} \\
\hline
%
$\max |\eta(e/\mu)|$ &  3 & 2.5 \\
$\min p_T(e/\mu)$ & 10 GeV & 15 GeV \\ \cline{2-3}
$\min \Delta R (b,e/\mu)$ &  \multicolumn{2}{c||}{0.6} \\
$\max m_T(e/\mu,\nu)$ & \multicolumn{2}{c||}{80 GeV} \\ \hline
Veto efficiency for jet/$e$ with &  \multicolumn{2}{c||}{} \\
15 GeV $< p_T(\hbox{jet}) $, $|\eta(\hbox{jet})| < 4.5$ &
 \multicolumn{2}{c||} {100\%} \\ \hline
Veto efficiency for jet/$e$ with &  & \\
10 GeV$ < p_T(\hbox{jet}) < 15$ GeV, $|\eta(\hbox{jet})| < 4.5$
&100\% & 50\% \\ \hline
\end{tabular}
\vspace{.2in}

{\small
Table 1. Acceptance and cut parameters, based on the ATLAS detector, for
two Scenarios: (i)~optimistic, and (ii)~conservative.}

\pagebreak
\begin{tabular}{||c|r|r|r|r||} \hline
            &  $\mH $ &  $\mH$  &  $\mH$   &  $\mH $ \\
  Processes &  90 GeV & 100 GeV &  110 GeV & 120 GeV \\ \hline
$WH$   &  155.6 & 116.6 & 86.8 & 59.7  \\
$Wjj$  &     101.6  & 90.4 & 75.3 & 58.1 \\
$Wbb$  &  198.7     & 154.7 & 122.7  & 98.3  \\
$WZ$   &  165.8     & 84.2 & 8.7 & 2.5 \\ \hline
$tt$   &  & & & \\
150 GeV  &   116.1 & 113.4 & 106.8 & 101.0 \\
175 GeV  &   45.3  & 45.1 & 45.7 & 46.5 \\
200 GeV  &   17.5  & 19.0 & 20.2 & 20.5 \\ \hline
$tbq$   &  & & & \\
150 GeV  &   75.5 &    68.3 &   59.4 & 49.6 \\
175 GeV  &   63.4 &   59.3  &  53.2  &   47.1 \\
200 GeV  &   52.2 &    53.0 &   46.9 &    45.2 \\ \hline
$tb$     &    &  &  &  \\
150 GeV  &  75.0 & 70.7 & 70.2 & 66.4 \\
175 GeV  &  34.8 & 38.9 & 38.5 & 36.5 \\
200 GeV  &   17.8 & 19.7 & 21.1 & 19.9 \\ \hline
$tq$     &       &	&	&   \\
150 GeV  &    35.8 &    36.9 &   36.9 &    37.7 \\
175 GeV  &   23.8 &    25.3 &   26.6 &    27.9 \\
200 GeV  &   15.3 &    17.4 &   18.4 &    19.6 \\ \hline
\end{tabular}
\vskip .1in
{\small
Table 2. Event rates for optimistic Scenario~(i) (see Table~1) at the LHC,
assuming 10 fb$^{-1}$ of yearly integrated luminosity. For the top
quark backgrounds, the top quark mass begins each respective row.
}
\vskip .1in
\hskip -.5in
\begin{tabular}{||ll||r|r|r|r||} \hline
  &  &$\mH $ &  $\mH$  &  $\mH$   & $\mH $ \\
\multicolumn{2}{||c||}{Scenario (i) High Acceptance} &90 GeV & 100 GeV &  110

GeV & 120 GeV \\ \hline
$\mt=150$ GeV: & S/B &   155.6/768.5 & 116.6/618.6 & 86.8/480.0 & 59.7/413.6 \\
& Significance &  5.6 & 4.7 & 4.0 & 2.9 \\
& yrs. for $5\sigma$ detection &  0.8 & 1.1
& 1.6 & 2.9 \\ \hline
$\mt=175$ GeV: & S/B &   155.6/633.4 & 116.6/497.9 & 86.8/370.7 & 59.7/316.9 \\
& Significance &  6.2 & 5.2 & 4.5 & 3.4 \\
& yrs.  for $5\sigma$ detection &  0.7 &
0.9 & 1.2 & 2.2 \\ \hline
$\mt=200$ GeV: & S/B &   155.6/568.9  & 116.6/438.4 & 86.8/313.3 & 59.7/264.1
 \\
& Significance &  6.5 & 5.6 & 4.9 & 3.7 \\
& yrs. for $5\sigma$ detection &  0.6 & 0.8
& 1.0 & 1.9 \\ \hline
\end{tabular}
\vskip .1in
{\small
Table 3. Scenario (i): Signal statistical significance and years required for
$5\sigma$ detection, assuming assuming 10 fb$^{-1}$ of yearly
integrated luminosity.
}

\pagebreak
\begin{tabular}{||c|r|r|r|r||} \hline
            &  $\mH $ &  $\mH$  &  $\mH$   &  $\mH $ \\
  Processes &  90 GeV & 100 GeV &  110 GeV & 120 GeV \\ \hline
$WH$ &  101.5 & 83.9 & 66.7 & 48.2 \\
$Wjj$  &  44.8 &   42.9 &   41.3 &    37.6 \\
$Wbb$  &  119.6 & 104.5 & 89.7 & 75.3 \\
$WZ$   &  97.1 & 52.0 & 5.8  & 1.8 \\ \hline
$tt$    &  & & & \\
150 GeV  &   84.2 & 87.2  & 90.0 & 84.2 \\
175 GeV  &   34.0 & 38.5 & 38.0  & 37.6 \\
200 GeV  &   14.0 & 16.7 & 17.5  & 17.8  \\ \hline
$tbq$      &  & & & \\
150 GeV  &   49.3    & 53.2  & 46.8    &  43.1 \\
175 GeV  &     38.2  &   42.8 &   42.9 &  41.8 \\
200 GeV  &   29.8    &   34.7 &  36.6  &  37.6 \\ \hline
$tb$     &    &  &  &  \\
150 GeV  &  49.6 & 52.0 & 53.7 & 50.8 \\
175 GeV  &   23.4 & 26.0 &  28.6 & 28.9 \\
200 GeV  &   12.6 & 14.5  & 16.0 & 16.6 \\ \hline
$tq$     &       &	&	&   \\
150 GeV  &   21.2 &    24.3 &   25.9 &    26.2 \\
175 GeV  &   13.9 &    16.4 &   18.2 &    19.9 \\
200 GeV  &    8.8 &    10.7 &   12.6 &    13.9 \\  \hline
\end{tabular}

\vskip .1in
{\small
Table 4. Event rates for the more conservative Scenario~(ii) (see
Table~1) at the LHC, assuming 10 fb$^{-1}$ of yearly integrated luminosity.
For  the top quark backgrounds, the top quark mass begins each respective row.
}

\vskip .1in
\hskip -.5in
\begin{tabular}{||ll||r|r|r|r||} \hline
  &  &$\mH $ &  $\mH$  &  $\mH$   & $\mH $ \\
\multicolumn{2}{||c||}{Scenario (ii) Conservative} &90 GeV & 100 GeV &  110 GeV

& 120 GeV \\ \hline
$\mt=150$ GeV: & S/B &   101.5/465.8 & 83.9/416.1 & 66.7/353.2 & 48.2/319 \\
& Significance &  4.7 & 4.1 & 3.5 & 2.7 \\
& yrs. for $5\sigma$ detection &  1.1 & 1.5 & 2.0 & 3.4 \\ \hline
$\mt=175$ GeV: & S/B &   101.5/371 & 83.9/323.1 & 66.7/264.5 & 48.2/242.9 \\
& Significance &  5.3 & 4.7 & 4.1 & 3.1 \\
& yrs.  for $5\sigma$ detection &  0.9 & 1.1 & 1.5 & 2.6 \\ \hline
$\mt=200$ GeV: & S/B &   101.5/326.7 & 83.9/276 & 66.7/219.5 & 48.2/200.6 \\
& Significance &  5.6 & 5.1 & 4.5 & 3.4 \\
& yrs. for $5\sigma$ detection &  0.8 & 1.0 & 1.2 & 2.2 \\ \hline
\end{tabular}
\vspace{.2in}

{\small
Table 5. Scenario (ii): Signal statistical significance and years required for
$5\sigma$ detection, assuming assuming 10 fb$^{-1}$ of yearly
integrated luminosity.
}
\pagebreak

\vskip .1in
\hskip -.5in
\begin{tabular}{||c||r|r|r|r||} \hline
 $S/B$-enhancing cuts imposed  &$pp \to WH+X$ &  $pp \to W+X$  &  $pp \to t+X$

&
$S/B$ \\ \hline

No additional cuts  & 83.9 & 205.0   & 123.7  & 0.26 \\

 $\dzphibb > 0.5 $  & 14.0 &  19.0   &  18.5  & 0.37 \\

 $|\zcmHc | <  0.4$   & 48.9 &  73.0   &  71.1  & 0.34 \\

 $|\zcmHc | <  0.4$,  $M_{Wb\bar b} >  230$~GeV  &
	                29.3 &  31.4   &  48.3  & 0.37 \\

 $\zcmbb > 0.3 $  & 13.0 &  21.7 &  4.5  & 0.50 \\

 $\zcmbb > 0.3$, $|\zcmHc | <  0.5$, & 10.8 &   9.2 &  2.5  & 0.92 \\
  and $M_{Wb\bar b} >  230$~GeV  & & & & \\ \hline
\end{tabular}
\vspace{.2in}

{\small
Table 6. The effect of $S/B$-enhancing cuts on the signal and the background,
for $\mt = 175$~GeV, $\mH = 100$~GeV, in events per year (assuming 10
fb$^{-1}$ of yearly integrated luminosity).
}

\newpage

\begin{figcap}

%
%

\item
{\small
The differential cross section $d\sigma/d\dzphibb$ versus
 $\dzphibb$  for $\mt = 175$~GeV and $\mH = 100$~GeV at the LHC,
for the signal (solid curve), the sum of $W$
backgrounds (dashed curve), and the sum of top quark backgrounds
(dotted curve).  The acceptance cuts are according to scenario (ii).
}

\vspace{.25 in}

\item
{\small
The differential cross section $d\sigma / d\zcmH$ versus
$\zcmH$ in the CM frame of $W+b\bar b$  for $\mt = 175$~GeV and
$\mH = 100$~GeV at the LHC, for the signal (solid curve), the sum of
 $W$ backgrounds (dashed curve), and the sum of
top quark backgrounds (dotted curve).
 The acceptance cuts are according to scenario (ii).
}

\vspace{.25 in}

\item
{\small
 The differential cross section $d\sigma/d \zlabH$ versus
$\zlabH$ in the laboratory frame  for $\mt = 175$~GeV and
$\mH = 100$~GeV at the LHC, for the signal (solid curve), the sum of $W$
backgrounds (dashed curve), and the sum of
top quark backgrounds (dotted curve).
The acceptance cuts are according to scenario (ii).
}

\vspace{.25 in}

\item
{\small
 The differential cross section $d\sigma/d\zcmHc$ versus
$\zcmHc$ in the reconstructed CM frame of $W+b\bar b$ for $\mt = 175$~GeV
and $\mH = 100$~GeV at the LHC, for the signal (solid curve), the
sum of $W$ backgrounds (dashed curve), and the sum of
top quark backgrounds (dotted curve).
The acceptance cuts are according to scenario (ii).
Here the CM frame was
reconstructed using the $W$ boson solution nearest (in $\Delta R$) to
the reconstructed Higgs boson.
}

\vspace{.25 in}

\item
{\small
  The differential cross section $d\sigma/d\zcmbb$ versus
$\zcmbb$ in the CM frame of $W+b\bar b$ for $\mt = 175$~GeV and
$\mH = 100$~GeV at the LHC, for the signal (solid curve), the sum of $W$
backgrounds (dashed curve), and the sum of top quark backgrounds
(dotted curve).
The acceptance cuts are according to scenario (ii).
Here the CM frame was
reconstructed using the $W$ boson solution nearest (in $\Delta R$) to
the reconstructed Higgs boson.
}

\vspace{.25 in}

\item
{\small
The differential cross section $d\sigma/d ({\rm min} p_T(b))$
versus the smaller of the  transverse momenta of the two bottom quarks
for $\mt = 175$~GeV and $\mH = 100$~GeV at the LHC, for the signal (solid
curve), the sum of $W$ backgrounds (dashed curve), and the sum of top quark
backgrounds (dotted curve).
The acceptance cuts are according to scenario (ii).
}

\vspace{.25 in}

\item
{\small
The differential cross section $d\sigma/d({\rm min}E(b))$ versus
the smaller of the energies of the two bottom quarks for $\mt =  175$~GeV
and $\mH = 100$~GeV, for the signal (solid curve), the sum of $W$
backgrounds (dashed curve), and the sum of
top quark backgrounds (dotted curve).
The acceptance cuts are according to scenario (ii).
}

\end{figcap}

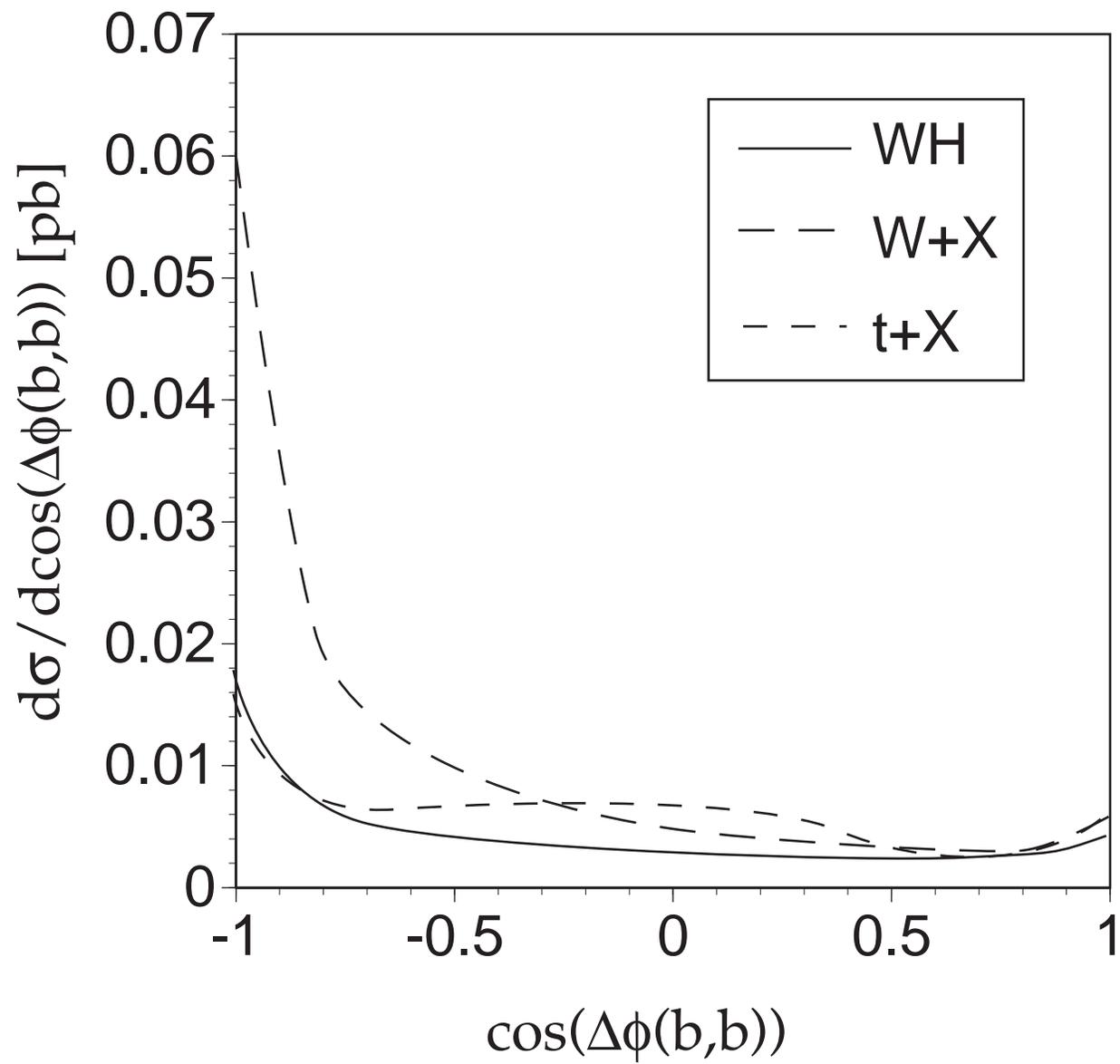
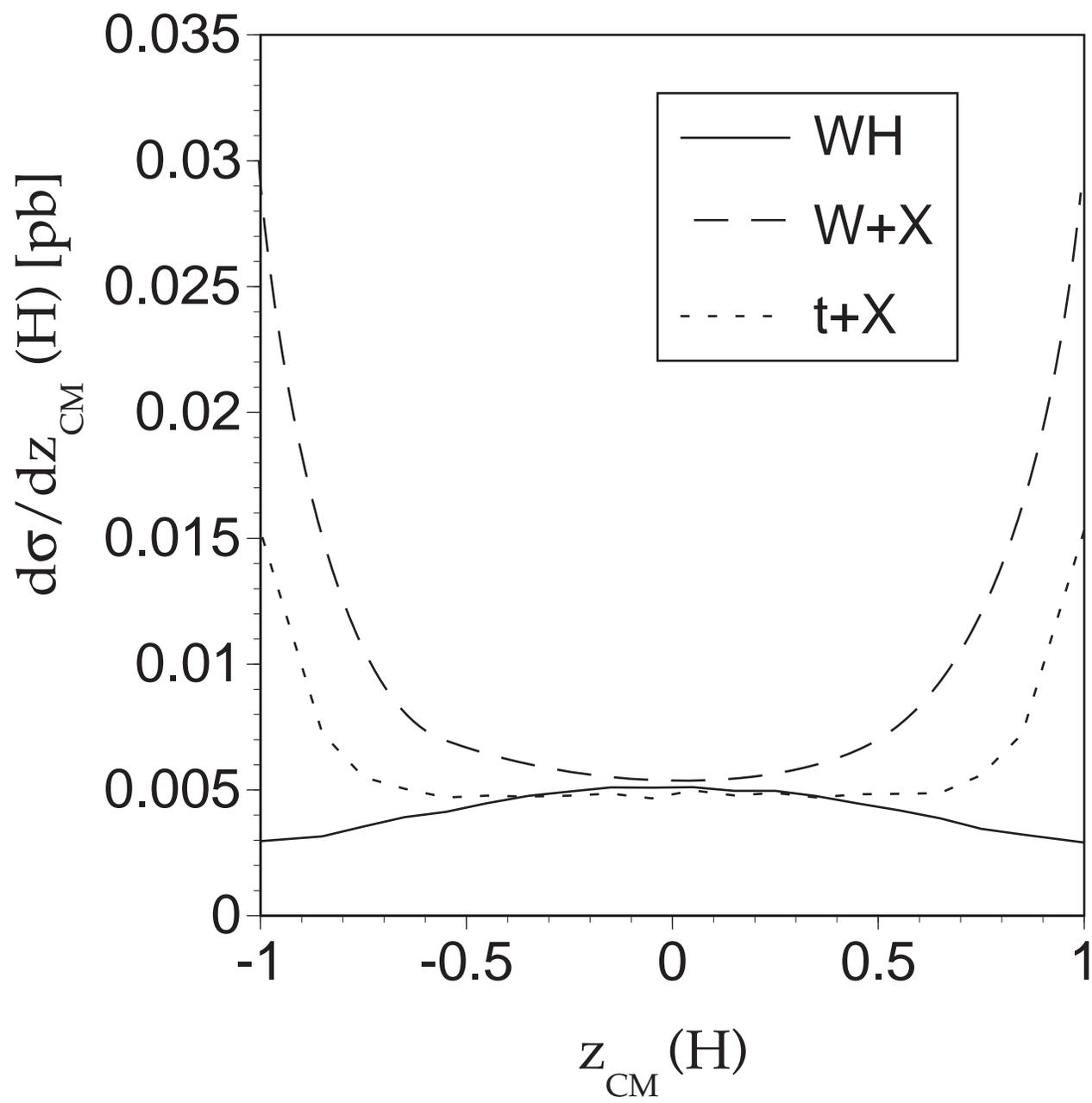
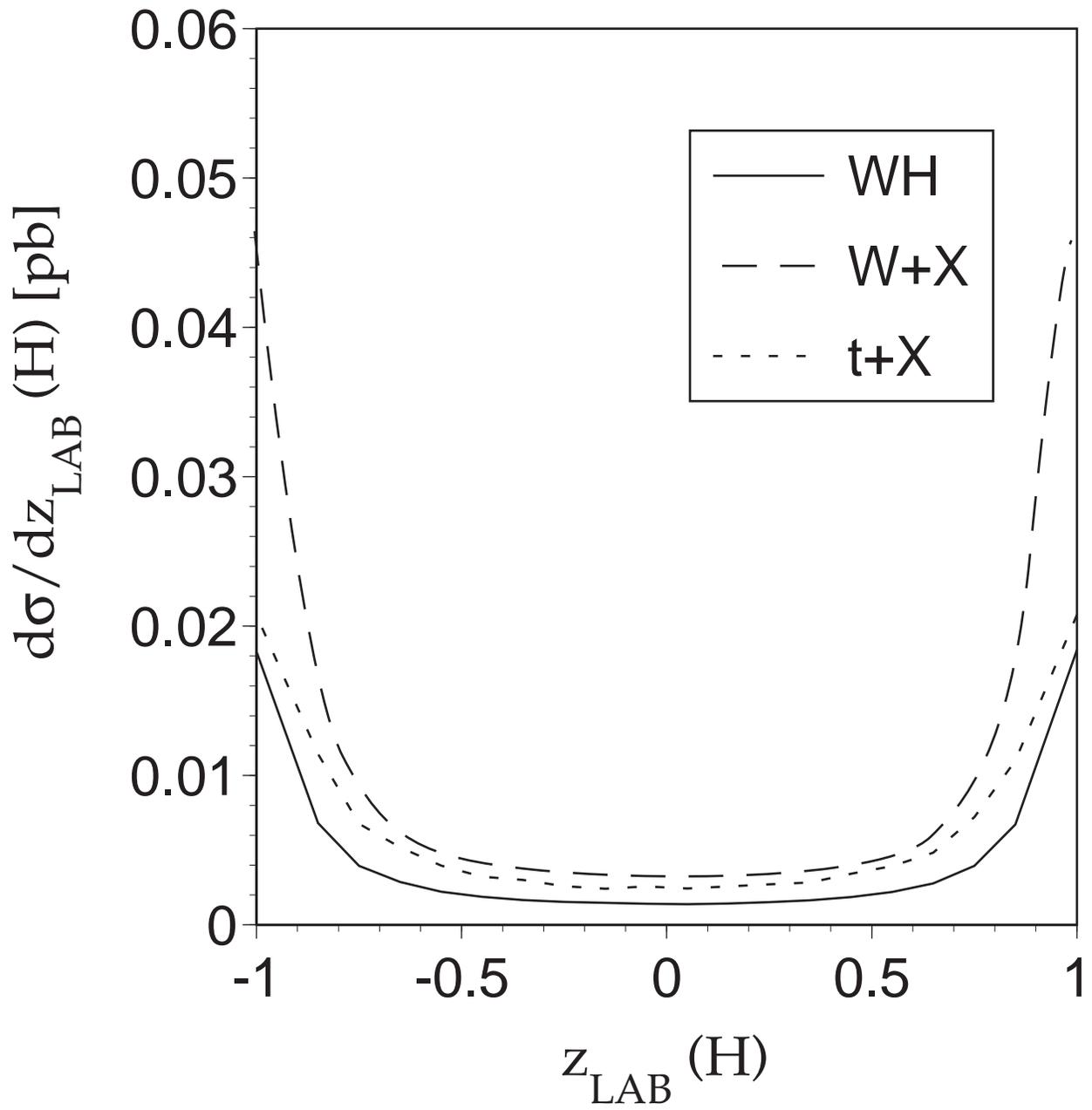
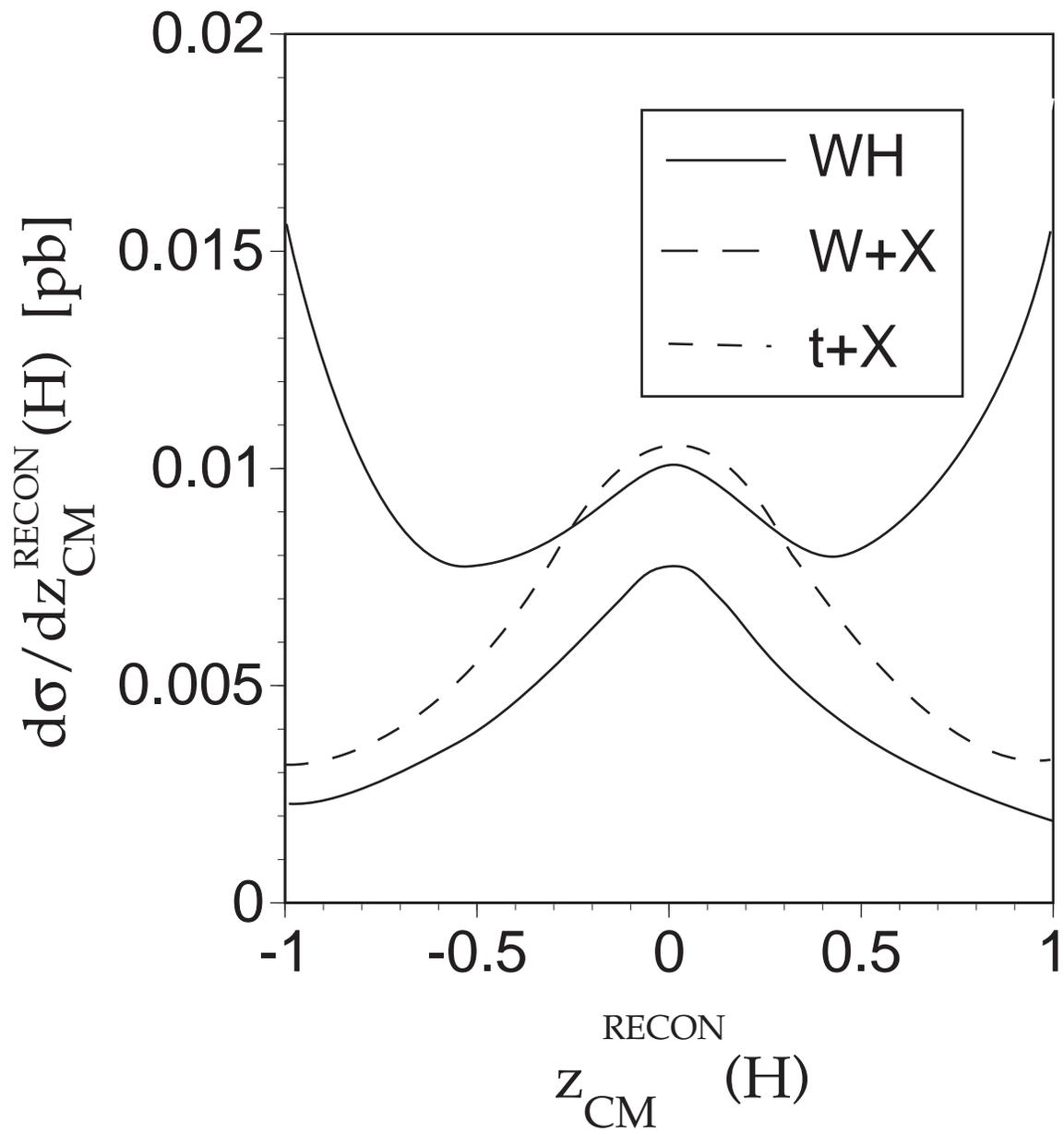
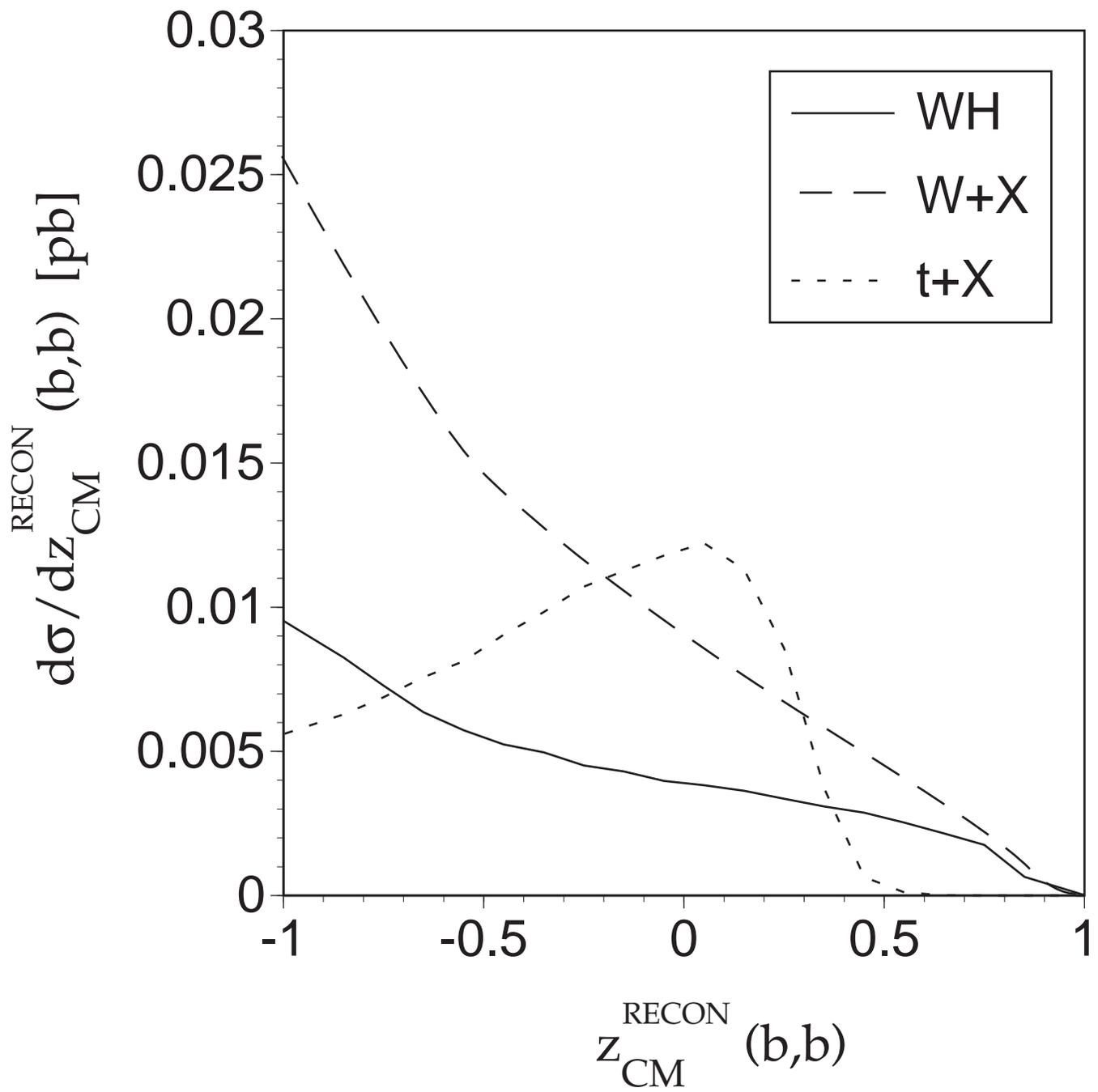
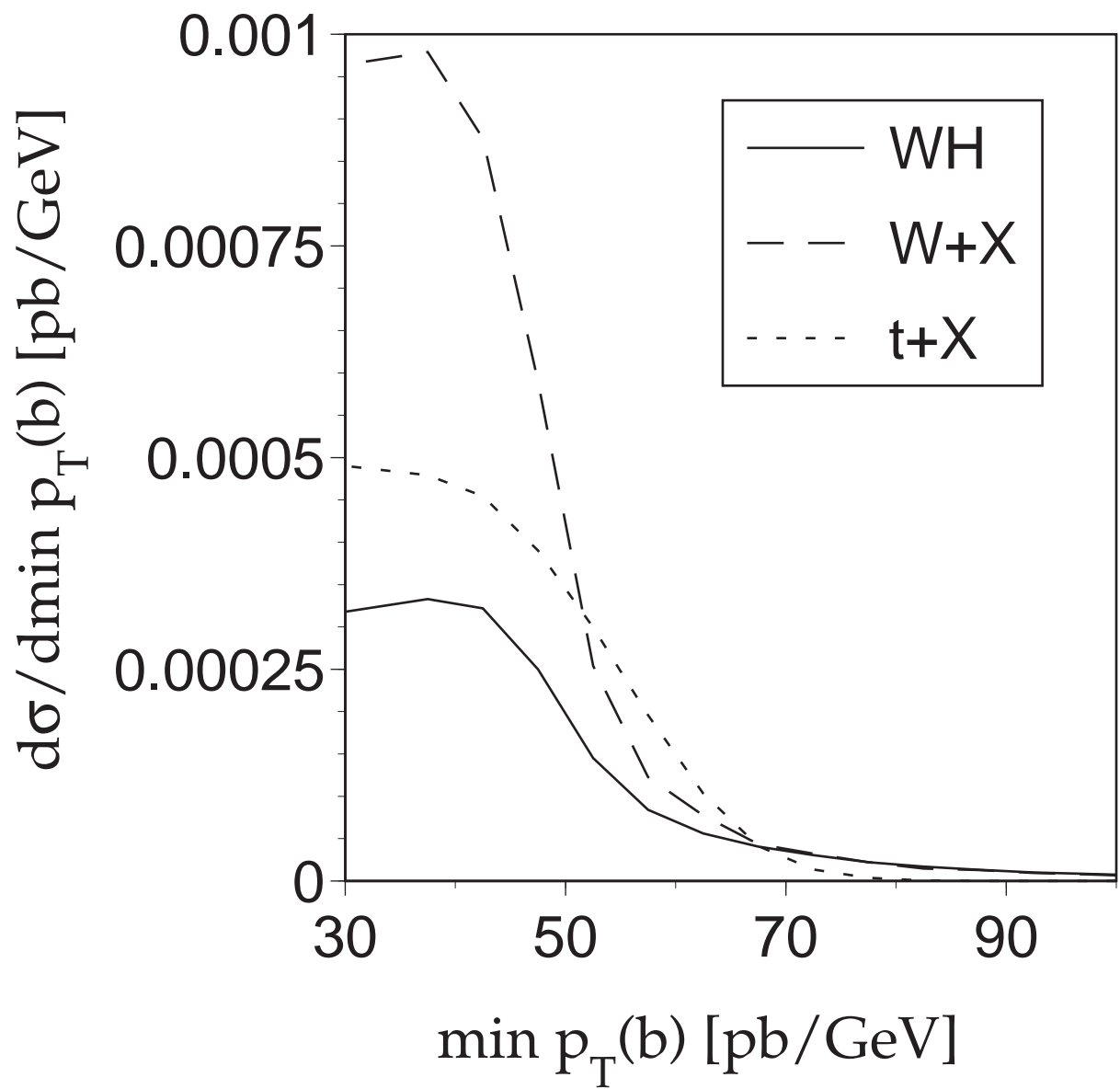
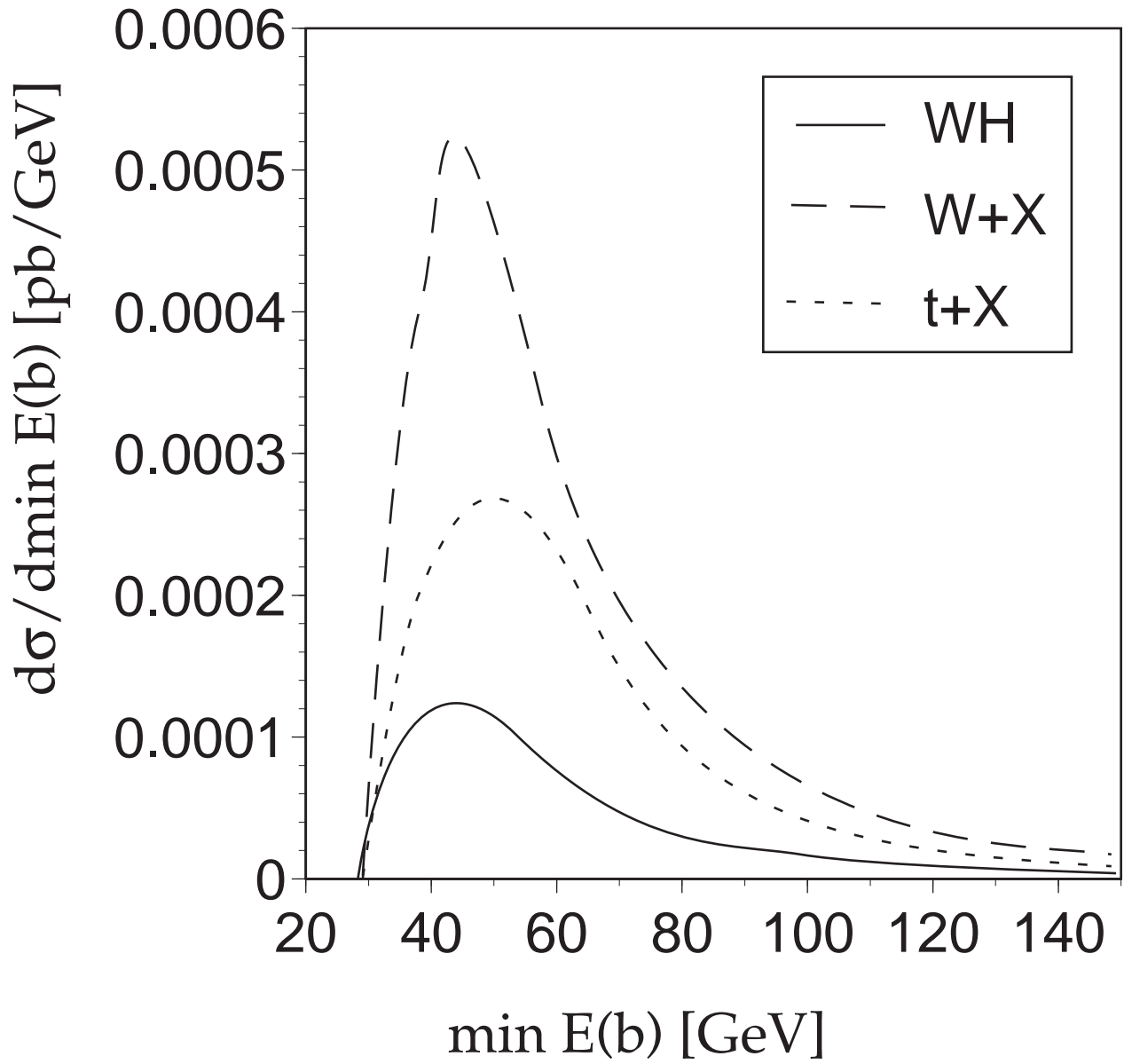
\end{document}